\documentclass[12pt]{article}
\usepackage[dvips]{graphicx}
\usepackage{epstopdf}

\title{Sampling the density of states}

\author{Marco Guagnelli\thanks{\texttt{marco.guagnelli@pv.infn.it}}\\
\small INFN, Sezione di Pavia\\[-0.8ex]
}

\date{}

\newcommand*{\Ne}{N_{\rm e}}
\newcommand*{\Nt}{N_{\rm t}}
\newcommand*{\Ns}{N_{\rm s}}
\newcommand*{\Nr}{N_{\rm run}}
\newcommand*{\deltaE}{\delta E}
\newcommand*{\Estar}{E^{*}}

\begin{document}
\maketitle

\begin{abstract}
It is shown that the algorithm introduced in \cite{langfeld} and conceived to deal with continuous degrees of freedom models is well suited to compute the density of states in models with a discrete energy spectrum too. The $q=10$ $D=2$ Potts model is considered as a test case, and it is shown that using the Maxwell construction the interface free energy can be obtained, in the thermodynamic limit, with a good degree of accuracy.
\end{abstract}

\section{Introduction}

In recent years Wang--Landau sampling (WLS) \cite{wl1,wl2} has became a standard tool in the area of Monte Carlo investigation of statistical systems and has been successfully applied to an ever increasing number of models and situations. Since its introduction, a lot of studies have been carried over in order to investigate its convergence properties and to refine it \cite{hklee,zhou1}.

The WLS finds its natural application to discrete energy models and its use for the study of continuous energy models deserves some care, since it has been found, for example, that a naive ``bin-discretization'' of a continuous energy leads to difficulties that need to be handled with care \cite{zhou2,xu,sinha}.

In a recent publication \cite{langfeld} a variation on the WLS theme, conceived for models with a continuous energy spectrum and based on a local linearization of the logarithm of the density of states (DOS), has been proposed. The purpose of this paper is to demonstrate, at numerical level, that the algorithm proposed in \cite{langfeld}, which will be referred to as the LLR sampling, or LLRS, is effective for models with a discrete energy spectrum too. There are in fact no {\em a priori} reasons the algorithm proposed in \cite{langfeld} should not work for discrete energy models, at least in the infinite volume (or thermodynamic) limit, in which the energy density, and accordingly the DOS, becomes in any case continuous. Clearly, on general ground, at finite volume discrepancies with respect to a standard WLS, at the level of ${\cal O}(1/L^{\alpha})$, where $\alpha$ depends on the specific observable and $L$ is the linear size of the lattice, are expected, but it will be shown that for $L$ large enough these finite size effects are small and do not affect a safe extrapolation of physical quantities to the infinite volume limit.

The motivation for applying the LLRS to discrete energy models stems from the fact that in some cases the derivative of the logarithm of the DOS respect to the energy ({\em i.e.} the inverse of the microcanonical equilibrium temperature at fixed energy) is all is needed: in such cases the LLRS gives the answer in a direct way and the overall computational cost can be dramatically decreased respect to WLS.

In the following the algorithm described in \cite{langfeld} will be introduced. Then as a first test the Ising model in $D=2$ will be addressed and the algorithm will be used to numerically extrapolate the microcanonical equilibrium temperature for a given energy to the thermodynamic limit; the comparison with the analytically known result will show that the LLRS does not produce systematic errors in the infinite volume limit. The third step will be the study of the interface free energy in the $D=2$, $q=10$ Potts model by means of the so called Maxwell construction, with the help of the WLS. Afterwards the same computation will be repeated with the LLRS, using larger lattices. At the end some conclusions will be drawn.

\section{The algorithm}

Generally speaking, the canonical partition function of a statistical system can be expressed as an integral (or a sum, for a discrete energy model) over all allowed energies:
\begin{equation}
{\cal Z}(\beta) = \int {\rm d}E g(E) e^{-\beta E},
\end{equation}
where $\beta \equiv 1/T$ is the inverse temperature and $g(E)$ is the DOS (the Boltzmann normalisation factor is set to 1 through all this paper). It is observed that in general, maybe neglecting the boundaries of the domain of definition, the logarithm of the DOS ({\em i.e.} the microcanonical entropy) is a smooth function of the energy (equivalently of the energy density); so smooth, in fact, that becomes meaningful to locally apply a linear approximation and to write
\begin{equation}
\label{eq:linapp}
\log g(E) = a(E_{0})[E-E_{0}] + c(E_{0}).
\end{equation}
It is understood that this approximation works only if $E$ is in the neighbourhood of $E_{0}$; to introduce the notation it can be written that eq. (\ref{eq:linapp}) works well for $E_{0}-\deltaE/2 \le E \le E_{0}+\deltaE/2$, with a suitable choose of $\deltaE$. 

The general strategy outlined in \cite{langfeld} will be now briefly recalled. Assuming the energy is restricted to the interval $E_{0}-\deltaE/2 \le E \le E_{0}+\deltaE/2$, the truncated expectation value of a general observable $f(E)$ can be defined:
\begin{equation}
\label{eq:principal1}
\langle\langle f(E_{0}) \rangle\rangle(a) = \frac{1}{\cal N}\int_{E_{0}-\deltaE/2}^{E_{0}+\deltaE/2} \textrm{d}E f(E) g(E) e^{-a E},
\end{equation}
where the normalisation ${\cal N}$ is given by
\begin{equation}
\label{eq:principal2}
{\cal N} = \int_{E_{0}-\deltaE/2}^{E_{0}+\deltaE/2} \textrm{d}E g(E) e^{-a E}.
\end{equation}
If eq. (\ref{eq:linapp}) was a perfect approximation then $a$ in eq. (\ref{eq:principal1}) could be chosen in order to cancel exactly $g(E)$; in this case a flat distribution would remain, so that
\begin{equation}
\langle\langle E\rangle\rangle(a) = E_{0} \qquad \textrm{for}\quad a=a(E_{0}).
\end{equation}
Assuming now that a good guess for $a(E_{0})$ (call it $a_{n}$) is known, ignoring higher order corrections and defining $\Estar\equiv E-E_{0}$, the truncated expectation value of $\Estar$ is readily obtained:
\begin{equation}
\label{eq:approx}
\langle\langle \Estar\rangle\rangle(a_{n}) = \frac{(\deltaE)^{2}}{12}[a(E_{0})-a_{n}]
+ {\cal O}(x^{3}\deltaE),
\end{equation}
where $x\equiv [a(E_{0})-a_{n}]\deltaE$. Eq (\ref{eq:approx}) can now be solved for $a(E_{0})$ in order to obtain a better approximation $a_{n+1}$:
\begin{equation}
\label{eq:iteration}
a_{n+1} = a_{n} + \frac{12}{(\deltaE)^{2}}\langle\langle \Estar\rangle\rangle(a_{n});
\end{equation}
the game can in principle be iterated up to convergence. 

Obtaining the truncated expectation value $\langle\langle \Estar\rangle\rangle(a_{n})$ is quite easy: all is needed is an entropic sampling \cite{lee}, restricted to the relevant energy range, with $g(E)$ given by eq. (\ref{eq:linapp}) and $a(E_{0})$ replaced by $a_{n}$.

As a matter of fact, as it is noted in \cite{langfeld}, it is quite unuseful to obtain a really precise estimate of $\langle\langle \Estar\rangle\rangle(a_{n})$ in order to increase the precision on $a_{n+1}$. In any case the value of $a_{n}$, after a suitable thermalisation period, will start to fluctuate around its mean value, but the amplitude of the fluctuations ({\em i.e.} the variance of $a$) is of course a function of the number of entropic sampling sweeps used to compute $\langle\langle \Estar\rangle\rangle(a_{n})$. A good compromise consists in using a reasonable number of entropic sampling sweeps $\Ne$, and to average $a_{n}$ over $1 \le n \le \Ns$ after $\Nt$ thermalisation steps. 

The above line of reasoning is enough to define the following algorithm to compute 
$\partial\log g(E)/\partial E$ for a given value of $E_{0}$ and a given choice of $\deltaE$:
\begin{enumerate}
\item guess a reasonable starting value for $a_{n}$ ($n = 0$); 
\item compute $\langle\langle \Estar\rangle\rangle(a_{n})$ performing $\Ne$ sweeps of entropic sampling with $\log g(E) = a_{n}E$ (the sampling is restricted to the range $[E_{0}-\deltaE/2,E_{0}+\deltaE/2]$; configurations whose energy is out of this range are simply not taken into account in the averaging process);
\item compute $a_{n+1}$ by using eq. (\ref{eq:iteration});
\item $n \leftarrow n+1$ and iterate;
\item discard the first $\Nt$ values of $a_{n}$ and average the remaining $\Ns$ values to obtain an estimate of $a(E_{0})$, which can be directly interpreted as $(\partial\log g(E)/\partial E)_{E=E_{0}}$.
\end{enumerate}

\section{Ising model}

The Ising model in 2D is the first--choice testbed because of the existence of an analytic solution against which numerical results can be checked. In this case, however, the analytic result at finite volume \cite{beale} are of little help, since the LLR sampling as defined above will show finite size effects respect to the ``true'' finite size results for $g(E)$. This is by no means a problem, since the check will be done against results in the thermodynamic limit.

The hamiltonian of the Ising model is taken to be
\begin{equation}
\label{eq:hamising}
{\cal H} = \frac{1}{2}\sum_{\langle i,j\rangle}[1-\sigma_{i}\sigma_{j}],
\end{equation}
where the sum is over nearest--neighbours sites and $\sigma_{i} = \pm 1$.

The LLR sampling defined in the previous section was used in order to compute $a(E_{0})$, with $E_{0} = 0.5\;V$, at several lattice size $L$, in order to proceed to the infinite volume limit. In this limit $a(E_{0})$ should become $(\partial S(E)/\partial E)_{E=E_{0}}$, a quantity for which an analytic answer is of course known. 
\begin{table}[!hbp]
\begin{center}
\begin{tabular}{|c|c|c|} 
\hline 
$L$     & $N_{\rm run}$ & $a(E/V=0.5)$  \\
\hline 
$64$    & $50$      & $0.754679(7)^{*}$ \\
$72$    & $50$      & $0.755187(6)^{*}$ \\
$80$    & $37$      & $0.755549(7)^{*}$ \\
$96$    & $50$      & $0.756020(5)$     \\
$112$   & $40$      & $0.756299(4)$     \\
$128$   & $40$      & $0.756494(3)$     \\
$160$   & $20$      & $0.756716(4)$     \\
$192$   & $20$      & $0.756837(5)$     \\
$240$   & $30$      & $0.756933(2)$     \\
$320$   & $18$      & $0.757005(2)$ \\
extrap. & -         & $0.757105(2)$ \\
\hline 
\end{tabular}
\end{center}
\caption{Simulations parameters and results for Ising model. Values marked with an $^{*}$ have not been used in the extrapolation to the infinite volume limit.}
\label{tab:contising}
\end{table}
For each value of $L$ the following procedure has been followed:
\begin{itemize}
\item after $\Nt=100$ thermalisation steps, $a$ has been averaged over $\Ns=1000$ measures, using $\deltaE=L$;
\item each measure, see eq. (\ref{eq:iteration}), has been obtained with $\Ne=1000$ entropic sampling sweeps, where one entropic sampling sweep consists in trying for $V=L^{2}$ times to flip a randomly chosen spin;
\item The procedure has been repeated $\Nr$ times in order to compute errors. Lattice sizes, $\Nr$ and results are shown in table \ref{tab:contising}.
\end{itemize}
A few comments about the parameters used in the procedure delineated above are in order:
\begin{itemize}
\item Starting with $a_{0}=0.8$, the convergence to the equilibrium value of $a$ is really fast and $\Nt=100$ is more than enough for all lattice sizes considered;
\item $\Ne$ is in any case taken proportional to the volume of the system in order to ensure that in the entropic sampling process each spin of the system has a chance to be flipped many times; it has to be taken into account that the ``flipping acceptance'' (in the present case in which $E/V = 0.5$), {\em i.e.} the probability that  the flip of a given spin during the entropic sampling process is accepted, is about $0.37$;
\item $\deltaE$ is taken proportional to $L$ in order to ensure that in the infinite volume limit this range do not shrink to zero when measured in ``physical'' units; the relevant point is that $\deltaE/V$ will go to 0 as $1/L$. Taking $\deltaE = L$ seems to be the most simple choice, and with this choice the linearization (\ref{eq:linapp}) is satisfied to an high degree of accuracy.
\end{itemize}
Moreover it has to be noted that a discrete energy model requires sums to be written instead of integrals in eqs. (\ref{eq:principal1}) and (\ref{eq:principal2}); accordingly, for the Ising model eq. (\ref{eq:iteration}) is modified as follows:
\begin{equation}
\label{eq:iterising}
a_{n+1} = a_{n} + \frac{12}{4(\deltaE) + (\deltaE)^{2}}\langle\langle \Estar\rangle\rangle(a_{n});
\end{equation}
\begin{figure}[ht]
\includegraphics[width=\textwidth]{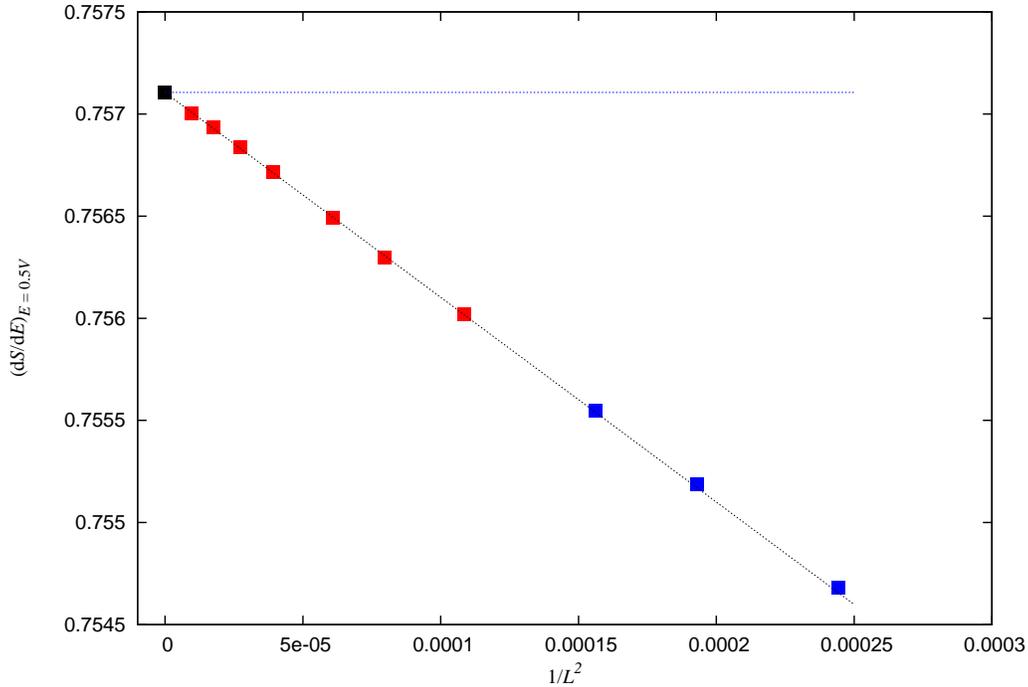} 
\caption{$(\partial S(E)/\partial E)$ infinite volume extrapolation for the Ising model. Fitting function to the data: $f(1/L^{2}) = f_{0} + f_{2}/L^{2}$.}
\label{fig:contising}
\end{figure}
Several lattices, ranging in size from $L=64$ up to $L=320$ have been considered, and in figure \ref{fig:contising} the thermodynamic extrapolation in $1/V$ (not considering the three smallest lattices, but their inclusion hardly changes the result) is shown. The final result, 
\begin{equation}
a(E_{0}) = 0.757105(2),
\end{equation}
has to be compared with the analytic answer, for which the convention defined in (\ref{eq:hamising}) gives, up to six digits, the value $0.757106$.
The almost perfect agreement shows that in the thermodynamic limit there are no systematic effects, at least in the high statistical precision reached. The extrapolation shows that the finite size effect on $(\partial S(E)/\partial E)$ for a given value of $E$ is proportional to $1/L^{2}$. The whole computation took about three days on a desktop computer (quadcore).

\section{The Maxwell construction}

In this section the $q=10$ Potts model in $D=2$, a model which is known to show a strong first order phase transition, will be considered; the hamiltonian of the model is given by
\begin{equation}
\label{eq:ham}
H = \sum_{\langle i,j\rangle} [1 - \delta_{\sigma_{i},\sigma_{j}}],
\end{equation}
where the sum is over nearest--neighbours sites and $\sigma_{i}$ is a Potts spin, taking values $1\dots 10$. The minimum of the energy is $0$ (ferromagnetic ground state) and the maximum energy attainable by the system is equal to the number of unsatisfied bonds, $2\;V$. 

\begin{figure}
\includegraphics[width=\textwidth]{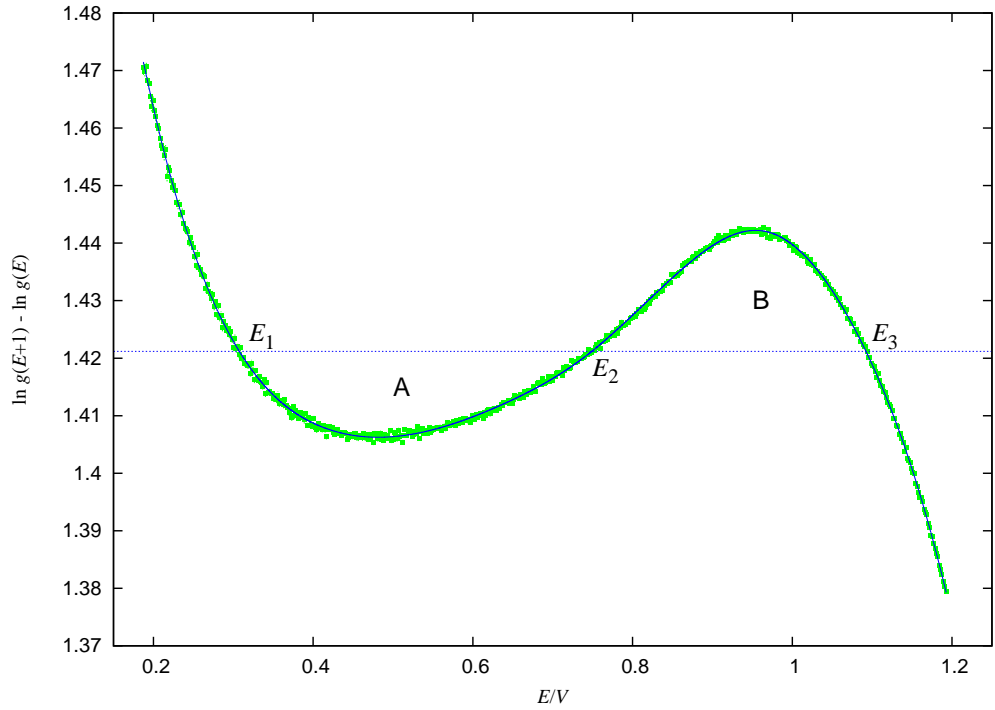} 
\caption{$\partial\log g(E)/\partial E$ for $D=2$, $q=10$ Potts model ($L = 24$). Green band: WLS results. Full blue curve: B-spline fit (12 free parameters) to data points. Dotted blue line: the critical value of $\beta\equiv 1/T$ as defined by the Maxwell construction (see text for the meaning of the labels).}
\label{fig:maxwell24}
\end{figure}

The Maxwell construction will be elucidated with the help of figure \ref{fig:maxwell24}.
With a standard WLS the DOS of the model for $L=24$ in the relevant energy range was built. The $1/\sqrt{\ln f}$ criterium \cite{zhou1} has been used to stop the updating of $\ln g(E)$ at each level of the algorithm, halted when $f$ reached the value $f_{\rm min} \simeq 1+10^{-7}$, and $2\times 10^{3}$ independent runs have been used to build the mean. The inverse of the microcanonical equilibrium temperature at energy $E$ was then estimated by
\begin{equation}
\beta(E) \equiv \frac{\partial S(E)}{\partial E} = \ln g(E+1)-\ln g(E).
\end{equation}
The result is shown as a green band in figure \ref{fig:maxwell24}.

When the two regions $A$ and $B$ have equal areas then the horizontal line which defines them at the same time defines the critical value of $\beta \equiv 1/T$ at which the first order phase transition occurs. For the two regions to have the same area the following condition has to be satisfied:
\begin{equation}
\label{eq:maxwell}
-\int_{E_{1}}^{E_{2}}\frac{{\textrm d}\log g(E)}{{\textrm d} E}{\textrm d} E + \beta(E_{2}-E_{1}) =
\int_{E_{2}}^{E_{3}}\frac{{\textrm d}\log g(E)}{{\textrm d} E}{\textrm d} E - \beta(E_{3}-E_{2}).
\end{equation}
This in turns leads to the condition that the double maxima in the energy probability distribution,
\begin{equation}
P(E) = \log g(E) - \beta E,
\end{equation}
take the same value. $E_{1}$ and $E_{2}$ identify the locations of the two maxima and $E_{3}$ is the point of the minimum in between them. It can be shown that the common area of the two regions $A$ and $B$ is directly connected to the interface free energy between the two phases \cite{berg1,borgs}.

In order to perform the integral, a cubic B-spline fit has been used to reconstruct the data, obtaining $F_{WL}(L=24) = 0.10600(31)$. The error has been computed by a standard jackknife procedure, and it has been checked that the value obtained is almost completely insensitive to the number of free parameters in the B-spline fit; going from $7$ to $13$ breakpoints ({\em i.e.} from $9$ to $15$ coefficients) the change in $F$ is very well below the statistical error.

A subtle point here is the choice of breakpoints locations. The procedure used is the one outlined in \cite{geneticspline}: instead of choosing equispaced breakpoints, the choice of their locations is demanded to a genetic algorithm, and the best set of breakpoints is determined by asking for the best overall $\chi^{2}$.

\begin{figure}
\includegraphics[width=\textwidth]{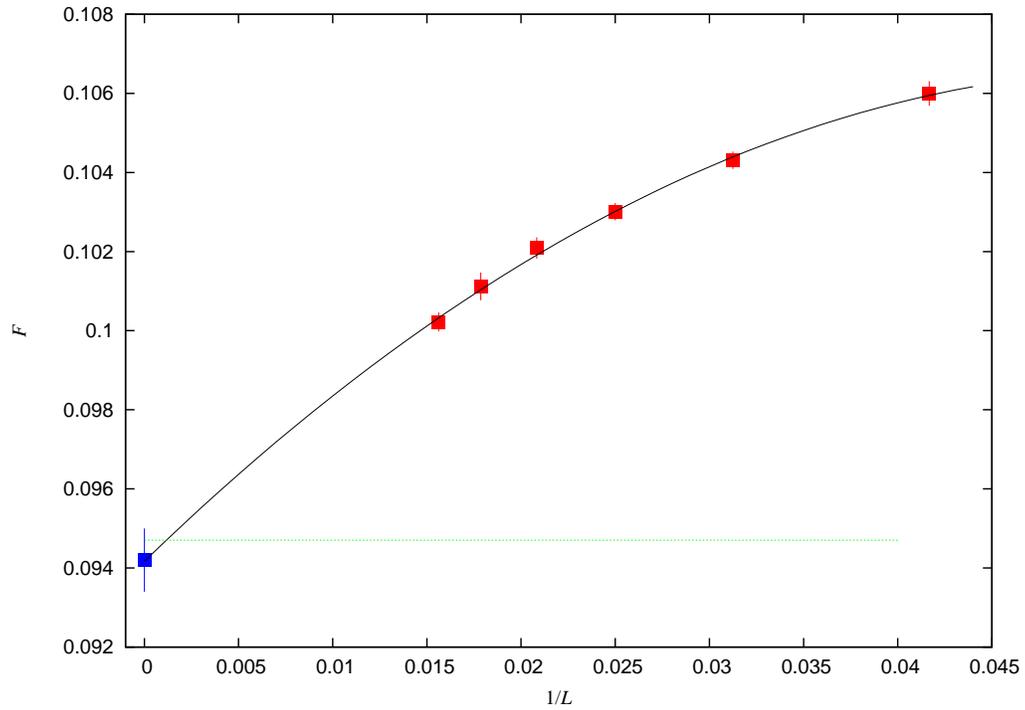} 
\caption{Extrapolation to the thermodynamic limit of the interface free energy for the  $D=2$, $q=10$ Potts model. Data (red squares) have been obtained with a standard WLS and the use of the Maxwell construction. The dotted green line represents the analytic result from \cite{borgs}.}
\label{fig:fwl}
\end{figure}

The same exercise has then been repeated for several lattice sizes (see table \ref{tab:fwl} for details) up to $L=64$. In figure \ref{fig:fwl} the extrapolation to the infinite volume limit is shown.
\begin{table}[!hbp]
\begin{center}
\begin{tabular}{|c|c|c|} 
\hline 
$L$ & $N_{\rm run}$ & $F$ \\
\hline 
$24$ & $2.0\times 10^{3}$ & $0.10600(31)$ \\
$32$ & $2.0\times 10^{3}$ & $0.10431(22)$ \\
$40$ & $1.8\times 10^{3}$ & $0.10301(22)$ \\
$48$ & $1.3\times 10^{3}$ & $0.10209(24)$ \\
$56$ & $3.8\times 10^{2}$ & $0.10112(35)$ \\
$64$ & $7.5\times 10^{2}$ & $0.10022(24)$ \\
\hline 
\end{tabular}
\end{center}
\caption{WLS simulation data for $D=2$, $q=10$ Potts model.}
\label{tab:fwl}
\end{table}
Data have been fitted with a quadratic function:
\begin{equation}
F(L) = F_{0} + F_{1}/L + F_{2}/L^{2},
\end{equation}
and the final result, $F_{0} = 0.0942(8)$ with $\chi^{2}/{\rm d.o.f.} \simeq 0.29$, compares quite well and is statistically compatible with the analytic result obtained in \cite{borgs}, $F_{0} = 0.094701$, and with the multicanonical result of \cite{billoire}.

\section{LLR sampling for the Potts model}

\begin{figure}
\includegraphics[width=\textwidth]{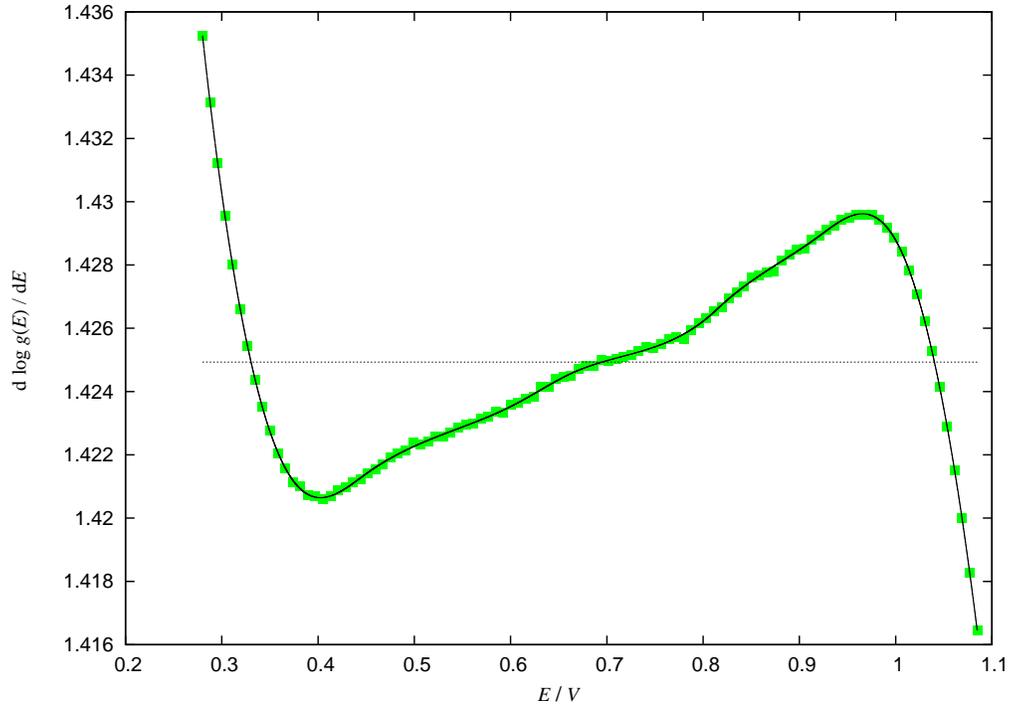} 
\caption{$\partial \log g(e)/\partial E$ for $L=128$. Continuous line is a cubic B-spline fit reconstruction of the data as described in the text. Dotted horizontal line is the value of $\beta_{c}$ computed via the Maxwell construction. Errors on data points are smaller--equal than symbols size.}
\label{fig:diff128}
\end{figure}

The aim of the previous section was only to show that the Maxwell construction, together with a good knowledge of the DOS, is a nice tool to numerically compute, in the thermodynamic limit, the interface free energy for spin models (specifically for the $D=2$, $q=10$ Potts model), and for this reason the WLS computation was not brought to large lattice sizes. In this section it will be described the LLRS computation of the same physical quantity. 
\begin{table}[!hbp]
\begin{center}
\begin{tabular}{|c|c|c|c|} 
\hline 
$L$ & $N_{\rm run}$ & $\beta_{\rm c}$ & $F$ \\
\hline 
$64$    & $310$ & $1.421595(8)$  & $0.11627(19)$ \\
$72$    & $242$ & $1.422550(9)$  & $0.11388(22)$ \\
$80$    & $176$ & $1.423222(9)$  & $0.11159(24)$ \\
$96$    & $183$ & $1.424084(7)$  & $0.10861(25)$ \\
$112$   & $149$ & $1.424582(6)$  & $0.10630(25)$ \\
$128$   & $136$ & $1.424932(6)$  & $0.10519(26)$ \\
$160$   & $119$ & $1.425359(5)$  & $0.10361(29)$ \\
$192$   & $79$  & $1.425555(9)$  & $0.10181(50)$ \\
extrap. &       & $1.426055(9)$  & $0.09501(54)$ \\ 
\hline 
\end{tabular}
\end{center}
\caption{LLRS simulation data and results for $D=2$, $q=10$ Potts model.}
\label{tab:f}
\end{table}
The procedure used is the following: the energy range from $E_{\rm min}/V\simeq 0.28$ to $E_{\rm max}/V\simeq 1.08$, which is the relevant one for studying the phase transition by means of the Maxwell construction, has been considered. The energy range has been divided in sub-intervals of size $L$ and for each sub-interval $a(E)$ (and whence $\partial\log g(E)/\partial E$) has been computed. The main difference respect to the WLS computation regards the fact that $\partial\log g(E)/\partial E$ is not computed for all values of $E$ but is only sampled, the sampling interval being $\Delta E = L$: the function is fully reconstructed at the end by means of the same B-spline fit procedure described above. The overall cost in terms of computer time is such that the extension of the computation of $F$ to larger lattice sizes is possible in a really reasonable amount of time. The function reconstruction can be judged in figure \ref{fig:diff128} (the case $L=128$ being taken as an example; for all other lattice sizes function reconstruction is of similar quality). Once $\partial\log g(E)/\partial E$ has been reconstructed, Maxwell construction can be used to obtain both the critical value of the temperature at finite volume and the value of the interface free energy. Note that the iterating equation for $a$ changes with respect to the Ising case and becomes
\begin{equation}
\label{eq:iterpotts}
a_{n+1} = a_{n} + \frac{12}{2(\deltaE) + (\deltaE)^{2}}\langle\langle \Estar\rangle\rangle(a_{n}).
\end{equation}
Several lattice sizes, from $L=64$ up to $L=192$, have been considered. Simulation parameters (and results for $\beta_{c}$ and $F$) are in table \ref{tab:f}. For each $L$, each run and each value of $E$, $a(E)$ has been computed by using $\Nt=100$, $\Ns=100$ and $\Ne=1000$. In each case the starting value of $a$ has been taken very close to the infinite volume value of $\beta_{c}$.
\begin{figure}
\includegraphics[width=\textwidth]{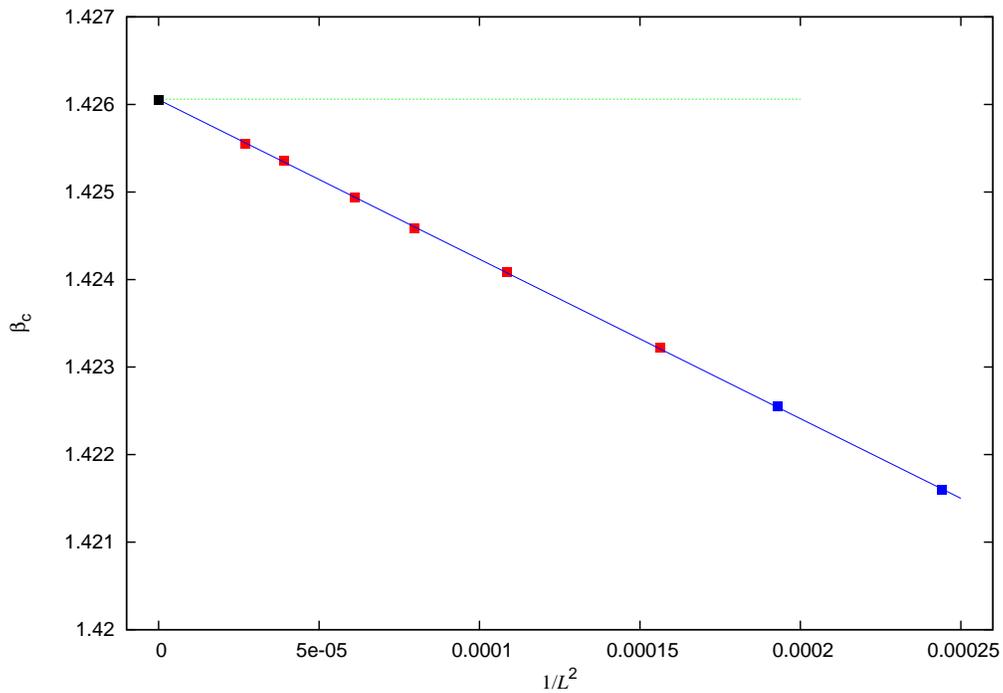} 
\caption{$\beta_{c}$ extrapolation to the infinite volume limit for the  $D=2$, $q=10$ Potts model. Data (red squares) have been obtained with LLR sampling and the use of the Maxwell construction. Blue points represent data not used in the extrapolation. The dotted green line represents the analytic known result $\beta_{c} = \log(1+\sqrt{10})$.}
\label{fig:betac}
\end{figure}
The first question to address is the computation of the critical temperature in the infinite volume limit. As can be seen in table \ref{tab:f} and figure \ref{fig:betac} is it possible to extrapolate the data to the thermodynamic limit as a linear function in $1/L^{2}$, obtaining a value for $\beta_{c} \equiv 1/T_{c}$ very close to the analytically known value: $\beta_{c} = \log(1+\sqrt{10})$. 
\begin{figure}
\includegraphics[width=\textwidth]{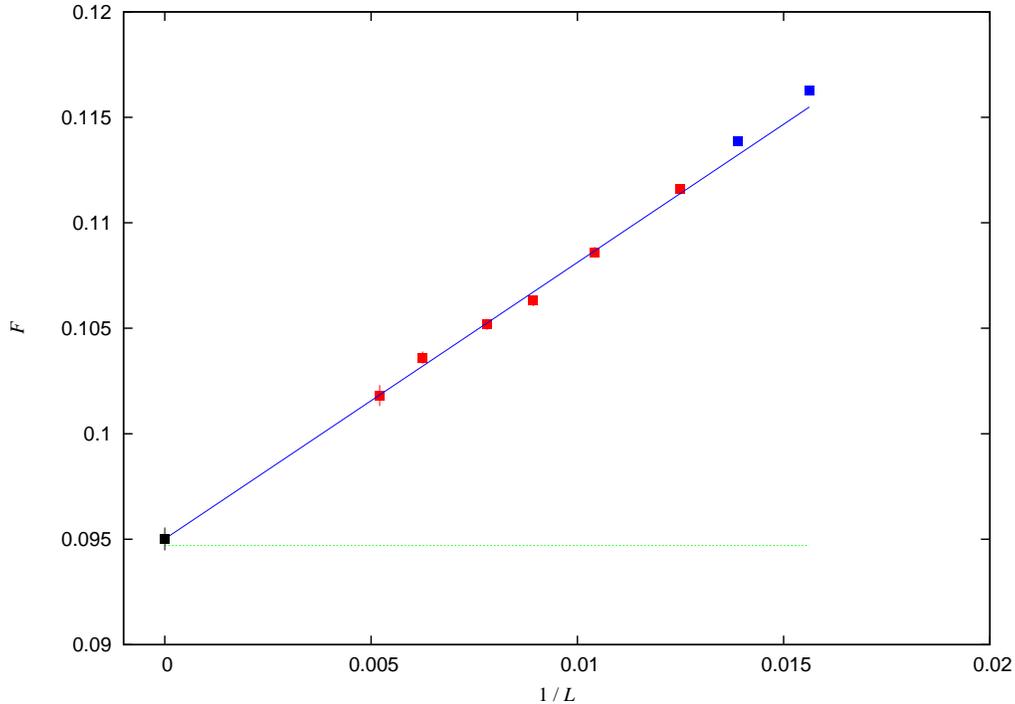} 
\caption{Extrapolation to the thermodynamic limit of the interface free energy for the  $D=2$, $q=10$ Potts model. Data (red squares) have been obtained with LLR sampling and the use of the Maxwell construction. Blue points represent data not used in the extrapolation. The dotted green line represents the analytic result from \cite{borgs}.}
\label{fig:f}
\end{figure}
In figure \ref{fig:f} the infinite volume extrapolation of the interface free energy is shown. Note that $1/L$ finite size effects are much bigger with respect to the WLS computation. Nevertheless, discarding the two smaller lattices and fitting linearly in $1/L$, a good $\chi^{2}$ around 1 can be obtained, and the final result is $F = 0.09501(54)$, which is again compatible with the analytic result of \cite{borgs} and with the multicanonical result of \cite{billoire}.

\section{Conclusion}

In this work it has been shown that also without a fine tuning of simulation parameters ($\Ne$, $\Nt$, $\Ns$, starting value of $a$, for example) the algorithm presented in \cite{langfeld}, which in this paper has been called LLRS, can be successfully used with discrete energy models, leading to results for physical quantities which in the statistical accuracy reached do not show systematic effects in the infinite volume limit. After a short exercise with the Ising model, the $D=2$ $q=10$ Potts model has been considered and the critical value of the temperature and of the interface free energy have been computed with a good degree of accuracy and a modest computational effort. The main motivation for using LLRS instead of WLS, in the case of discrete energy models, is the following: $\partial\log g(E)/\partial E$, which is the key quantity entering in the Maxwell construction, is directly computed and can be fully reconstructed for the whole relevant energy range by a careful sampling (what is meant here is that $\partial\log g(E)/\partial E$ can be computed for a sample of the discrete energy set and then safely interpolated). There is also to mention that in this way one can avoid all convergence problems of the energy histogram which have to be taken into account when dealing with WLS.

In this work the possibility to fully reconstruct the DOS, $g(E)$, starting from information extracted by LLRS, has not been explored. For continuous energy model a simple integration may suffice, and is limited only by discretisation effects. For discrete energy model, however, a modest extension of the procedure can allow for it. This will be the subject of a future work.

\section*{Acknowledgements} The author thanks B. Pasquini and S. Romano for a critical reading of the manuscript and for fruitful discussions. Part of the computation has been performed on CSN4 cluster.

\end{document}